\documentclass{sig-alternate}
\usepackage{amsmath}
\usepackage[pass]{geometry}
\usepackage{fancyhdr}
\usepackage{xcolor}
\usepackage{graphicx}

\usepackage[normalem]{ulem}
\usepackage[hyphens]{url}
\usepackage{microtype}
\usepackage{listings}
\usepackage[sort,nocompress]{cite}
\usepackage{soul}
\usepackage[inline]{enumitem}
\usepackage[font=small,labelfont=bf]{caption}

\newcommand{\ignore}[1]{}
\renewcommand{\tilde}[0]{$\sim$}

\newcommand{\hide}[1]{ }
\newcommand{\hints}[1]{ }

\newcommand{\myhl}[1]{#1}
\newcommand{\mysout}[1]{}

\newcommand{\ee}[0]{$^\dagger$}
\newcommand{\cs}[0]{$^\S$}

\fancypagestyle{firstpage}{
  \fancyhf{}
\setlength{\headheight}{50pt}

  \pagenumbering{arabic}
}

\title{A neural network memory prefetcher using semantic locality} 

	\author{Leeor Peled\ee ~~~~~~~~ Uri Weiser\ee ~~~~~~~~ Yoav Etsion\ee\cs\\
		\ee{}Electrical Engineering ~~~~~~~~~ \cs{}Computer Science \\
		Technion --- Israel Institute of Technology\\
		\normalsize\{leeor@tx, uri.weiser@ee, yetsion@tce\}.technion.ac.il}

\newcommand{\MAXGain}[0]{4.6$\times$\space}
\newcommand{\MaxSPECGain}[0]{2.7$\times$}
\newcommand{\AvgSPECGain}[0]{30\%\space}
\newcommand{\SPECGainOverGHB}[0]{43\%\space}
\newcommand{\SPECGainOverSMS}[0]{2$\times$\space}
\newcommand{\SPECGainOverVLDP}[0]{2.9$\times$\space}
\newcommand{\SPECGainOverRL}[0]{2.1$\times$\space}

\newcommand{\HistQ}[0]{128}

\newcommand{\numphases}[0]{4}

\usepackage[bookmarks=true,breaklinks=true,colorlinks,linkcolor=black,citecolor=blue,urlcolor=black]{hyperref}

\begin{document}
\maketitle
\thispagestyle{firstpage}
\pagestyle{plain}


\begin{abstract}

Accurate memory prefetching is paramount for processor performance, and modern processors employ various techniques to identify and prefetch different memory access patterns. 
While most modern prefetchers target \textit{spatio-temporal} patterns by matching memory addresses that are accessed in close proximity (either in space or time), the recently proposed concept of \textit{semantic locality} views locality as an artifact of the algorithmic level and searches for correlations between memory accesses and program state. 

While this approach was shown to be effective, capturing semantic locality requires significant associative learning capabilities.
In this paper we utilize neural networks for this task. 
We leverage recent advances in machine learning to propose a neural network prefetcher. 
We show that by \mysout{targeting semantic locality} \myhl{observing program context}, this prefetcher can learn distinct memory access patterns that cannot be covered by other state-of-the-art prefetchers. 

We evaluate the neural network prefetcher over SPEC2006, Graph500, and several microbenchmarks. We show that the prefetcher can deliver an average speedup of \AvgSPECGain\space for SPEC2006 (up to \MaxSPECGain) and up to \MAXGain over kernels. 
\mysout{We also explore the limitations of using neural networks for prefetching, and propose several power, energy and area optimizations.}
\myhl{We also present a high-level design of our prefetcher, explore the power, energy and area limitations, and propose several optimizations for feasibility.}
We believe that \mysout{future} \myhl{this line of research can further} improve the efficiency of such neural networks and allow harnessing them for additional micro-architectural predictions.

\end{abstract}

\section{Introduction}

Hiding memory access latency through memory prefetching is increasingly critical to processor performance. As a result, most modern processors employ multiple types of prefetchers to cover a wide range of application behaviors.
Existing prefetching schemes are designed to identify specific memory access patterns, ranging from sequential and strided access patterns to traversals over linked data structures. Most of these schemes target \emph{spatio-temporal locality}. They analyze the memory access stream and identify \emph{spatio-temporal relations} and instances of \emph{temporal correlation} between addresses or address-space artifacts (e.g., address deltas)~\cite{PrimerHWPref}.

An alternative view of data locality was recently proposed by Peled et al.~\cite{context_pref}. The proposed \textit{semantic locality} model argues that \emph{spatio-temporal locality} is a manifestation of a fundamental relation between accesses that is derived from the program semantics. According to this approach, access locality is an artifact of the algorithmic design of the code. Correlations between memory accesses may therefore be inferred from the program state, which in turn can be represented through various attributes of program and machine context at any point during the run.
However, the detection of state-address correlation is not deterministic. The simplistic reinforcement learning method proposed by Peled et al. extracted a number of attributes describing the program state and history, but could not discern between useful attributes and superfluous ones. This suggests that a stronger learning mechanism is required to extract semantic correlation.

Neural networks have become extremely popular in recent years as generic tools for solving complicated learning problems. Their structure allows them to solve non-linear classification problems, to learn patterns, and to perform associative reasoning tasks. As such, they appear to present a compelling alternative as the engine behind a semantic prefetcher. We believe that once technology advances enough to alleviate their power/performance costs, they will become the dominant predictors and prefetchers in future CPUs.  

In this paper, 
we present a neural network prefetcher that examines various machine and workload attributes and predicts memory addresses that will be accessed by the running program in the near future. 

Our exploration demonstrates that a neural network prefetcher can be dynamically trained to learn multiple access patterns that are today targeted by multiple, distinct, heuristic memory prefetchers.
We analyze the learning rate of neural networks and their applicability to dynamically learning memory access streams, the encoding of access patterns as synaptic weights in a network, and the network size required to store address correlations. In particular, we explore the limits of pattern complexity that various neural networks can learn, and examine their success over sequences representing the specialization of common existing prefetchers.
Alongside these benefits, we also study the technical challenges of implementing a neural network prefetcher. 

Our contributions in this paper are as follows:
\begin{itemize}
\item
  We present multiple methods for learning memory access patterns using neural networks. 

\item
  We explore the limits of pattern detection and predictability using different neural network layouts. We evaluate the size, depth and characteristics of networks suited for address prediction, and discuss various trade-offs in the implementation of a neural network prefetcher.

\item
  We define and implement a neural network prefetcher and evaluate its performance using gem5~\cite{gem5} over a variety of benchmark suites, including SPEC2006~\cite{spec2006}, Graph500~\cite{graph500}, and a variety of hand-written kernels. We show that our prefetcher can gain an average speedup of \AvgSPECGain over SPEC2006, and up to \MAXGain on some kernels, compared to a baseline without prefetching.
\end{itemize}

The remainder of this paper is organized as follows: Section~\ref{sec:learning} describes the challenges of contextual prefetching. Section~\ref{sec:nnpref} describes a context-based neural-network prefetcher. Section~\ref{sec:case_study} explores design considerations for the neural network.
Section~\ref{sec:related} presents related work. 
Section~\ref{sec:methodology} describes the experimental methodology used. 
Section~\ref{sec:evaluation} describes the simulation results, and we conclude in Section~\ref{sec:conclusions}.

\section{Learning memory access patterns}
\label{sec:learning}

Prefetchers observe the sequence of memory accesses performed by the processor and use that information to predict future memory accesses and fetch their associated data ahead of time.
The \textit{prefetch} problem statement can therefore be formulated as follows: at any point during the run, given the history of memory addresses previously accessed (and any complementary information available), produce the most likely N addresses to continue the sequence. 

When observing a sequence stripped of any additional information, this task is known as the problem of \textit{universal prediction}~\cite{UniversalPrediction}, which is prevalent in the field of information theory and often handled with probabilistic tools.
However, since we know that the stream represents accesses to structured data, constructed and used by an algorithm with some form of internal logic and recurrence, we have better chances of finding order in the underlying semantics that produced it.
 
\subsection{Identifying semantic locality}
Memory access streams are usually constructed of multiple sub-streams. Each of these sub-streams represents a portion of the application and is generated by distinct elements of the program semantics. The relations within such sub-streams can be defined as \textit{semantic locality}~\cite{context_pref} if they represent \textit{consequential} (not merely \textit{consecutive}) actions. Such relations may include pointer or index arithmetics, pointer dereferencing, or other forms of manipulations. 

The program interleaves these sub-streams, effectively forming a global stream that is further reordered at run-time by program control flow and sometimes also by an underlying out-of-order micro-architecture. This makes isolation of elements in the sequence very hard, since the prefetcher doesn't always know where to look for patterns, and since the patterns themselves often vary in their relation types.
\begin{figure}[t]
	\centering
	\includegraphics[width=1.2\linewidth]{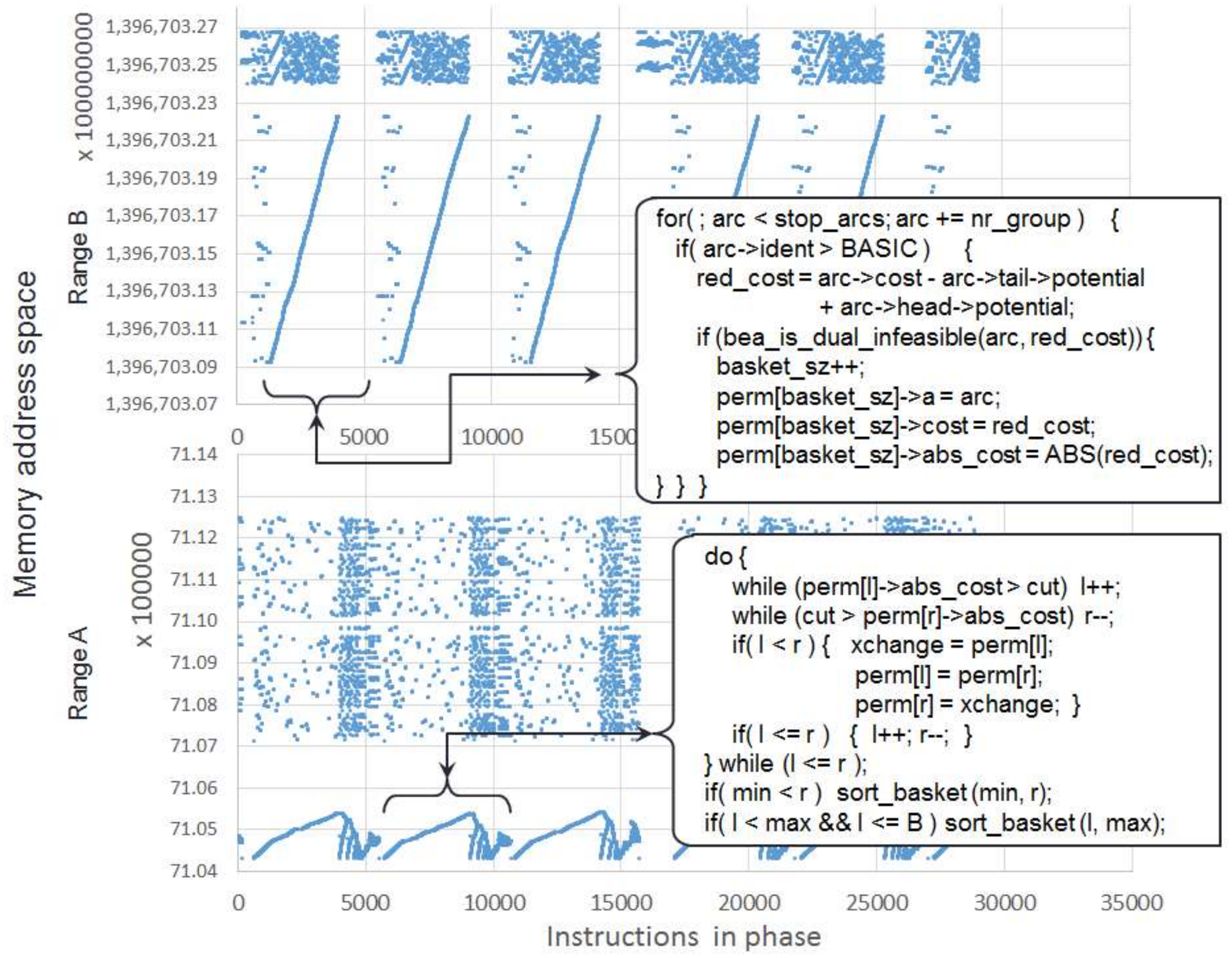}
	\caption{Access patterns during one of the phases of MCF, over several memory activity ranges (covering over 30\% of the runtime). The bottom range (also enlarged in Figure~\ref{fig:qsort}) exhibits an interesting pattern that represents a quicksort algorithm. Other patterns include direct linear strides representing the direct fields in the array of \textit{arc} structs, while the scatters represent indirectly linked elements accessed through the pointers there.}
	\label{fig:mcf_patterns}
\end{figure}
	
\begin{figure}[t]
	\centering
	\includegraphics[width=0.95\linewidth]{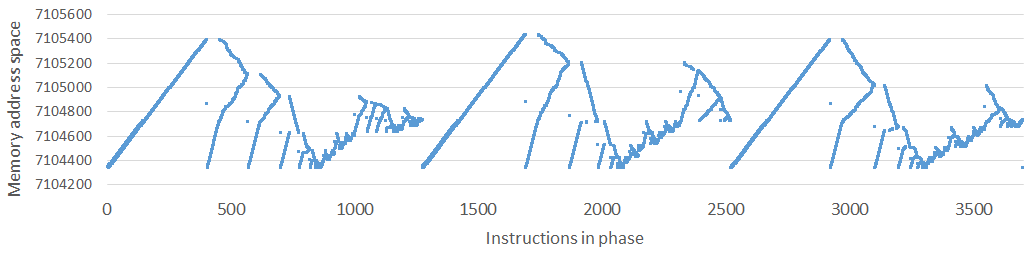}
	\caption{Zoom-in on several quicksort phases from Figure~\ref{fig:mcf_patterns}. The pivot in each step is chosen and elements are rearranged on its sides, creating a cone shape. Each pivot creates a skewed cone according to its location.}
	\label{fig:qsort}
	\vspace{-3mm}
\end{figure}

Figure~\ref{fig:mcf_patterns} shows an example of multiple sub-streams that are simultaneously active in the SPEC2006 MCF benchmark. The different memory ranges reflect different data sets (lists, trees, and arrays pointing at each other) that are interleaved by the algorithm. Each range has different characteristics, some with spatial layout and a linear access pattern, and others with more complicated accesses. For example, Figure~\ref{fig:qsort} illustrates a quicksort phase in MCF, going back and forth around a pivot element at each iteration. This mixture of access patterns demonstrates the interleaving problem, as each sub-stream would benefit from different prefetching techniques.

Most common prefetchers are designed to handle specific types of semantic relations and patterns, most often based on their manifestations as spatio-temporal locality or temporal correlation. Table~\ref{table:patterns} shows simple examples of such relations. Real applications with complex semantics often exhibit many types of relations (as shown in the MCF example), making the use of predefined relations insufficient.
\emph{One of our goals in this paper is to explore whether a sufficiently powerful and generic learning model can identify and learn all memory access patterns and semantic relations used by a program.}
\begin{table*}[t]
	\centering
	\scriptsize
	\begin{tabular}{l|c|c|c|c|c|c|c|p{3cm}}
		\textbf{Prefetcher} & \multicolumn{5}{c}{\textbf{Sample sequence}}&&  \textbf{Sample rule}&\textbf{Context info used} \\
		\hline
		\hline
		Streamer & A & A+1 & A+2 & A+3 & A+4 & A+5 & +1,+1,... & Last address \\
		\hline
		Strider & A & A+n & A+2n & A+3n & A+4n & A+5n & +n,+n,...& Last address \\
		\hline
	    SMS~\cite{SMS}& A & A+n & A+m & B & B+n & B+m & offsets: 0,n,m & Address delta history\\
		\hline
	    Markov~\cite{MarkovPredictors} & A & B & A & C & A & B & $A\xrightarrow{33\%}B $ ;  $ A\xrightarrow{66\%}C$ & Last address \\
		\hline
	    VLDP~\cite{VLDP}, GHB */DC~\cite{GHB} & A & A+2 & A+3 & A+5 & A+6 & A+8 & +2,+1,+2,+1..& Address delta history\\
		\hline
	    GHB PC/*~\cite{GHB} & $A_1$ & $B_1$ & $A_2$ & $B_2$ & $A_3$ & $B_3$ & $PC_A:{A}_n$ ;  $ PC_B:{B}_n$ & PC + Addr. delta history \\
		\hline
	\end{tabular}
	\vspace{3pt}
	\caption{Example pattern types targeted by common prefetchers, including streamers/striders, Spatial-based prefetchers (SMS) which repeat a pattern around anchor addresses, Markov prefetchers which rely on a probability model, VLDP and delta-correlation GHB which learn recurring delta patterns, and GHB/PC which records patterns based on PC.}
	\label{table:patterns}
	\vspace{-5mm}
\end{table*}

\subsection{Observing semantic locality by context}
\label{sec:context}
Since memory accesses are derived from the algorithms, the data structures, and the programming paradigms used, the program context itself can become a useful hint in the deobfuscation of the access stream. 
For that reason, most advanced prefetchers employ some portions of the program context to isolate substreams and discover patterns within them. The last column in Table\ref{table:patterns} shows examples of the context attributes used by each prefetcher. 
This raises the question whether truly abundant context will allow  better pattern isolation and detection of true semantic locality.

We define a \textit{program context-state} as a state vector with the current values of a set of attributes representing the CPU and program state (e.g., register values, branch and access history, access type). Each memory access has a distinct context-state at the time of dispatch that has some correlation to what the program was doing at that exact moment. Even though we cannot \textit{infer} what the program was doing from this information, we can still \textit{correlate} between that action and the state of the program. Strong correlation between past contextual states and future addresses often indicates true semantic relation~\cite{context_pref}. 



However, a context that is not detailed enough may fail to expose the desired relations, while an overly detailed context may overfit.
An effective context can only be found using the attributes that are semantically meaningful for a given relation (e.g., the register used for computing an address, or an IP for recurring accesses into a data structure). 
Selectively picking context attributes and identifying their relations to future memory accesses of the program require unique learning capabilities. 
Thanks to recent developments in the field of machine learning, we believe this can be achieved through online training of neural networks.


\section{The contextual neural network prefetcher}
\label{sec:nnpref}

The proposed neural network prefetcher predicts future memory accesses based on current program context.
We use a neural network to extract the contextual information and derive prefetch patterns that were associated with similar contexts in the past. 

\begin{figure}[t]
	\centering
	\includegraphics[width=0.95\columnwidth]{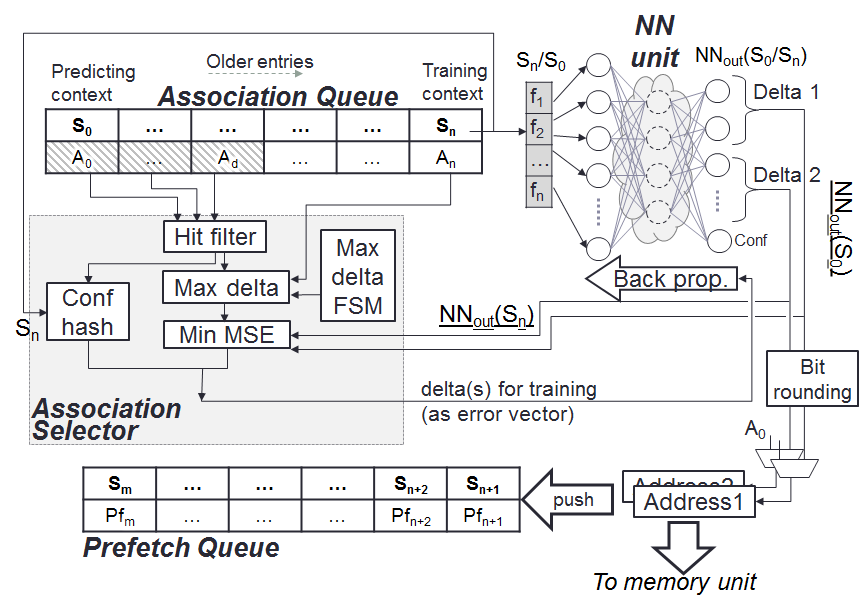}
	\caption{NN prefetcher diagram, showing the association queue, the NN unit, the logic for selecting which associations to train, and the prefetch queue that keeps prefetches for feedback.}
	\label{fig:NN_design}
\end{figure}

\begin{figure}[t]
	\centering
	\includegraphics[width=1\columnwidth]{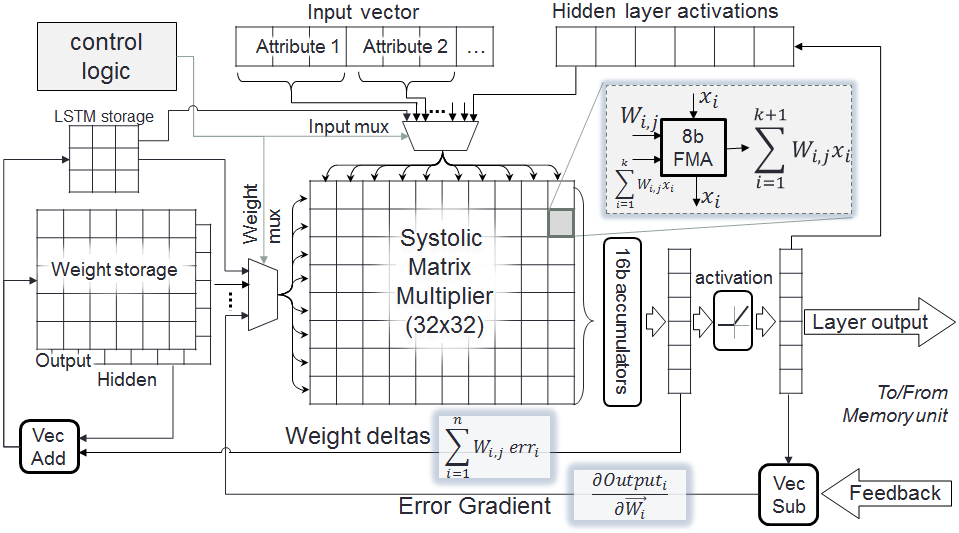}
	\caption{The "heart" of the NN prefetcher is a 32x32 systolic array matrix multiplier with 8b precision. The array is fed by various input/error buffers (depending on the phase) and the weight vectors, managed by a controller.}
	\label{fig:NN_design2}
\end{figure}

\begin{figure}[t]
	\centering
	\includegraphics[width=0.8\columnwidth]{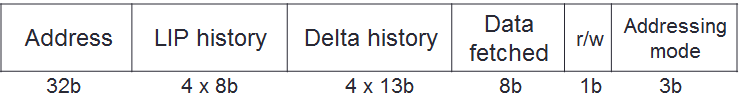}
	\vspace{2mm}
	\caption{128-bit context state vector used for semantic approximation. LIP history concatenates bits [8:1] of recent memory access LIPs, delta history concatenates bits [14:2] of the deltas between them.}
	\label{fig:context_vec}
\end{figure}

The proposed prefetcher, shown in Figures~\ref{fig:NN_design} and~\ref{fig:NN_design2}, receives the stream of memory addresses accessed by the program and the context state at the time of each access. The context states are represented as bit-vectors describing CPU attributes, as shown in Figure~\ref{fig:context_vec}. 
These attributes were already shown~\cite{context_pref} to be useful in semantic correlation of various access stream patterns. Elements representing accumulated history values are implemented using shift registers and concatenate a subset of each value.

\begin{figure}[t]
	\centering
	\includegraphics[width=0.9\columnwidth]{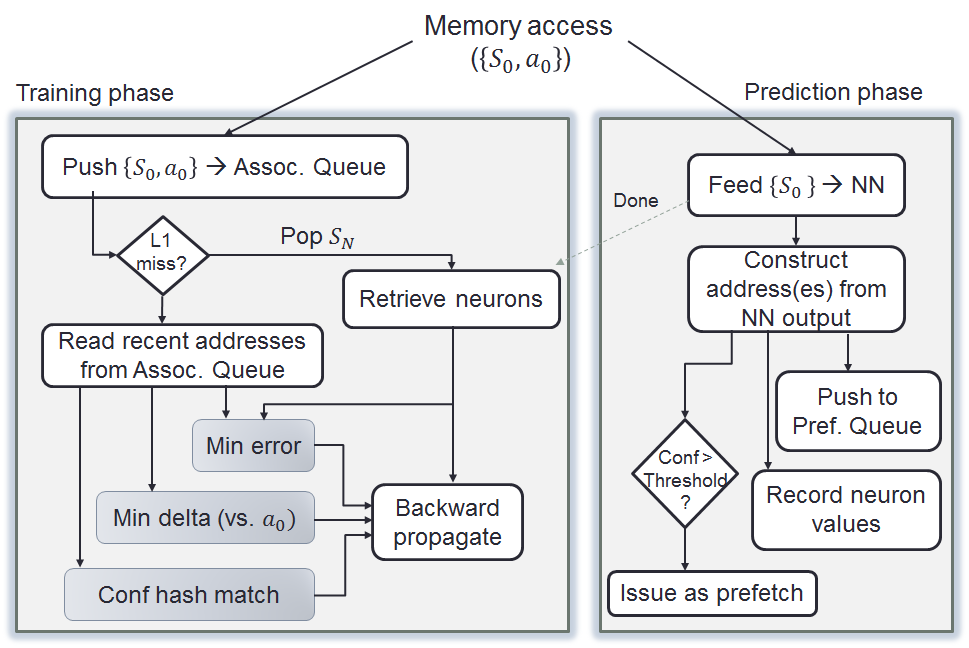}
	\caption{Schematic workflow of the neural network prefetcher given memory access to $a_0$ under context state $S_0$}
	\label{fig:NN_flow}
\end{figure}

The prefetcher is divided into two main parts: 
\begin{itemize}
	\item The association unit, shown in Figure~\ref{fig:NN_design}, is responsible for producing and tracking the context-address associations.
	\item The neural network unit, shown in more detail in Figure~\ref{fig:NN_design2}, is responsible for learning the associations and producing predictions based on context state vectors.
\end{itemize}
Figure~\ref{fig:NN_flow} describes the main workflow of the prefetcher.

\subsection{Selecting context-address associations}
\label{sec:association_queue}
The prefetcher must be able to train on unsupervised samples of context-address pairs based on the observed history. We do not know at the time of training whether the associations will prove to be semantically related. Thus, we rely on recurrence to strengthen true relations over time.

Since we want to establish a useful prefetch distance, we must represent a long window of history between each context and the associated address it should predict. This presents the challenge of picking context-address pairs within this window that are likely to be semantically related and are not too close or too far away. 
To address this challenge we maintain the \textit{association queue}. This queue stores the history of context states and addresses observed on every memory unit access.
On every access we push the current state vector and address at the head of the queue, and pop the element at the tail which represents the oldest stored context state $S_N$. We then choose one of the recent addresses from the queue to associate with it. Since the addresses are selected from around the head of the queue, they represent a distance of roughly $N$ accesses (where $N$, the queue size, is selected to fit the desired prefetch depth given the machine average miss time). We used a queue of \HistQ{} elements in our runs. 

Ideally, we would prefer a fixed distance between the associated context state and address, to better capture recurrence. However, the actual distance of semantically related pairs may change between iterations due to out-of-order execution or code path changes. To mitigate, we allow some variance in the association depth. 
We achieve that by observing the most recent addresses ($A_0$ .. $A_d$) as a possible association for $S_N$, selecting the best candidate for a match.  

\hide{In this paper we use a delta-prediction mode, where the output is trained to be the address delta relative to the address of the associated context ($A_N$). This allows bounding the stored values (usually 16 bits are enough), as well as providing better recurrence on strided patterns. However, unlike most delta based prefetchers, outs is not limited to small deltas within the same page; storing the deltas is therefore not inferior to storing the full addresses (aside from the need to support negative values). This is important when trying to achieve partial memoization of fractured linked data structures.}

Notably, all addresses that reported an L1 cache hit are filtered out of the selection process under the assumption that we should focus on predicting accesses that incur miss penalties. Nevertheless, they still allocate an entry in the association queue for correct depth tracking, and their context state may still trigger predictions. Accesses that hit a cache line installed by a prefetch are treated like misses so that we may strengthen the useful association they represent.

During the training phase, $S_N$ is first fed into the neural network to produce a prediction on the output. This output is matched against the recent $d$ addresses read from the queue, and the closest match is selected (the address $A_i$ such that the delta from the predicting address $A_i - A_N$ has the most bits matching the network output: $argmin_i(popcnt((A_i - A_N) \oplus NN_{out}))$). We rely on the slow momentum of the gradient descent method to produce the strongest recurring pattern for that given input and further strengthen such associations, assuming they represent a semantic connection. In parallel, this eliminates random or transient associations. 

Since address deltas are more likely to be associated within related memory ranges, it is best to restrict the \textit{maximal delta} we may associate.
This restriction is very useful in some cases, but may become detrimental if the data sets are highly fractured. We therefore apply a dynamic limit, where the maximal delta limit is incremented when the prefetcher is deemed insufficiently useful. If the percentage of useful prefetches (i.e., prefetches that are hit by demand accesses) stays below a predefined threshold over a period of time, we raise
the maximal delta limit (linearly, by multiples of 0x2000) to allow farther associations. These values were tuned to achieve a fast but thorough scan.
We allow several cyclic sweeps through possible maximal deltas, each having a chance to generate useful new associations that were forbidden before, and eventually select the value that performed best. 

In addition, we observed the need to kickstart the initial convergence of the neural network, especially in the presence of conflicting associations (occasions when the same context is associated with different addresses). For that purpose, we add a \textit{context hash}: each context vector being associated will store the associated delta into a table indexed by a hashing of the context vector. When selecting a candidate for association, we prefer addresses that already appear in the context hash for the given context. The context hash has smaller storage lifetime than the neural network (it will be overwritten on either conflicting associations or on overloaded hash values), 
but it gives us an additional level of confidence that the trained context/address pair was already observed, and it helps the neural network train faster over recurring associations by reducing the noise. 

%


\subsection{Parallel association policies}
\label{sec:association_policies}
As described in the previous section, there are several possible ways of selecting the desired candidate address for association with each context.
Different association policies match different types of patterns that the program may exhibit. Therefore, to extract all possible patterns, we can place multiple networks in parallel and train each one independently with the address associated by each policy. 
Since the inputs to the networks are all identical and some of the hidden neurons may be useful for different networks, we optimize this by dividing only the output layer. 
This way, different associations may be trained in parallel over a single network, allowing shared storage and shared learning.

In our experiments we used one subset range of nodes to train associations with minimal distance in terms of minimal square error (MSE) from the current output of the network. We use the other subset to train the best context-hash match (i.e., the first context-address pair that hits in the context hash). 
The hash-based associations often give better results during early stages of the run, as they are faster to train, while the slower MSE-based associations improve the overall coverage.
 
\hide{We attempted using up to four parallel networks, where the other two are trained over PC confidence (i.e., prefer associating addresses with highest recurring program counters), or simply choosing another best MSE match (since it is done on a separate network, it will learn different predictions over time and make different min-MSE associations). However, while the context-hash-based network was shown to be critical for performance, the other two had only marginal benefits.}

The predictions generated by the multiple output subsets are interpreted as deltas relative to the address of the predicting state. We construct the prefetch addresses and send them to the memory unit. 
Since triggering prefetches increases overall system bandwidth, the proposed prefetcher may dispatch \textit{"shadow prefetches"}, similar to the RL-context prefetcher~\cite{context_pref}. This is done by storing all generated prefetches into a \textit{prefetch queue} and using them to collect feedback, but dispatching only the highest confidence ones to become actual prefetches.

\subsection{The NN prediction unit}
\label{sec:NN_design}
The central component in our prefetcher is the neural network itself, described in Figure~\ref{fig:NN_design2}. It consists of 3 layers of fully connected \textit{perceptrons}. Each perceptron models a simplified \textit{neuron} by performing a linear product between its inputs and a weight vector, and then applying a non-linear activation function on the result (we use ReLU~\cite{ReLU} for quantized networks, and a logistic sigmoid~\cite{sigmoid} for full precision ones).
The depth, layout and precision of the neural network are explored in Section~\ref{sec:case_study}. Area/power mitigations are discussed in Section~\ref{sec:feasibility}.  


The prefetcher performs two steps for each L1 cache miss (triggering on average every~\tilde{10} instructions, based on an L1 hit rate of~\tilde{80\%}). First, the \textit{prediction} phase performs \textit{inference} over the most recent context state vector ($S_0$): it feeds the input context vector to the entry layer, computes the neuron activations at each layer, and interprets the values at the output layer as a binary representation of the predicted value (as an address delta relative to the predicting address $A_0$). If the result is distinct enough (the output bits are within valid thresholds), it is sent to the memory unit to be prefetched.

During inference, the input vector is 
fed into a 32 row $\times$ 32 column systolic array. Each column in the array represents an input element, and each row represents a single hidden neuron. Each matrix element multiplies the input value with the corresponding 8-bit floating point weight assigned to it by the neuron. The results are propagated to the next column and accumulated with a 16-bit accumulator per row. The controller then selects the next set of weights and issues another phase of operations for the next set of inputs, while accumulating the partial sums. Since the hidden layer has 32 elements in our design, the neurons can all be computed in parallel, and we require \numphases\space phases to compute the entire input vector. However, thanks to the systolic array, we can start propagating the next phase in a pipelined manner. 

Once the hidden layer is calculated, the results pass through an activation unit 
and are sent back to the input latch for the computation of the output layer. The controller fetches the output weights in parallel. Since there are 32 output neurons and 32 hidden neuron inputs, computing the output neurons requires 4 additional computation phases over the systolic array. The result is activated again and sent back to the main prefetcher unit to construct the address for prefetching. 
The neuron values for $S_0$ are also stored into the association queue.

Subsequently, the prefetcher begins the \textit{training} step by popping the oldest context state, $S_n$, from the association queue. The neuron values kept (from the time this context state was added to the queue and passed through the inference step) are preloaded into the matrix and we perform \textit{back-propagation} by comparing them with the desired value (chosen by each of the policies described in Section~\ref{sec:association_policies} and converted into address deltas relative to $A_n$).  We use gradient-descent to minimize the bitwise min square error (MSE) vector:  
\small
\begin{equation}
\delta_{W_{output}} = \frac{\partial MSE}{\partial W_{output}} = \frac{\partial MSE}{\partial output} * \frac{\partial output}{\partial prod} *
\frac{\partial prod}{\partial W_{output}}, 
\end{equation}
\normalsize

where $prod$ is the vector of pre-activation values, and $output$ is the output-layer neuron vector.
Since MSE is the squared distance, the derivative is proportional to the delta. Therefore, the first term is a simple subtraction. The second term is the derivative of the activation, which (for ReLU) is a simple step function around 0. The third term is the hidden neuron value. The overall calculation is therefore a 32-element subtraction (which is shared by all trained neurons) followed by an element-wise multiplication for each of the output weight matrix elements, and the result is added to the current weight and is updated in the weight matrix.

After the output weight matrix is updated, the hidden neuron weights are updated in a similar fashion. However, in the hidden layer the impact of each weight affects all output neurons. Therefore, the error gradient of each neuron must incorporate the errors of the entire output layer. This requires a preliminary pass to compute the weighted errors (one FMA per hidden layer weight), followed by another delta calculation phase.



\subsection{Prediction feedback}
The neural network must be able to accept feedback for training, both positive and negative. We achieve this by tracking predictions (both real ones and shadow prefetches) in the \textit{prefetch queue}. When a demand hits this queue at a depth deemed useful (we use the same weight function as in~\cite{context_pref}), we trigger another training pass on the NN to strengthen the context/address pair that was hit.
If, however, a prediction is hit outside the useful range, or if it drops off the queue without ever being hit, negative feedback must be provided. The neural network is less equipped to provide such feedback as there is no way to ``untrain" a sample. Instead, we remove a single bit from the NN outputs interpreted as predicted delta, and instead train it to hold the confidence of the prediction. Any positive feedback will train it to a high value, and any negative feedback will train it to a low one. Since, unlike the address bits, this output neuron is not intended to be interpreted as a binary value, we are not 
forced to train it as such. Instead, we can assign gradual values according to the strength of the feedback 
and achieve more accurate training.

\subsection{Information encoding}


Unlike other learning mechanisms that rely on direct or semi-associative storage, neural networks store their knowledge in the form of node weights. Rather than performing lookups in internal tables, the neural network calculates the outcome using multiple paths that combine the values of all the active nodes in parallel. 
Consequently, we do not control how information will be stored in the neural network, and the actual layout depends on the convergence of the training process. Notably, a single node in the network might sometimes be associated with a specific, highly specialized feature of the input pattern (sometimes known as the ``grandmother cell"~\cite{grandmotherCell} and referring to how single neurons can learn to recognize abstract concepts). 

This form of distributed storage increases the number of unique patterns a neural network can learn (i.e., the network's \textit{expressiveness}) compared to direct tabular representation.  
For example, a network that learns an \textit{add} function may first attempt to memorize all encountered results, but after sufficiently long training may eventually converge towards a simpler implementation of a bitwise adder simply because memoization will no longer fit in the network's storage capacity.

\textit{Recurrent neural networks} (RNNs) represent another important enhancement for neural networks. Normal feed-forward networks connect neurons only between consecutive layers and are effectively stateless with regard to the input stream. 
RNNs add the notion of loops within the layers, allowing the network to preserve memory of the previous learning steps. Previous inputs take part in the inference of the current input, allowing the network to learn \textit{temporal} functions. This makes such networks effective in learning sequences and patterns (as opposed to a set of unordered samples), since these often represent some forms of temporal relations.

In this work we focus on a recent type of RNN called \textit{Long Short-Term Memory} (LSTM)~\cite{LSTM2}. This variant, which adds neurons functioning as ``gates" to control the internal loops, is able to learn when the internal node should be used, adjusted, or reset. 
LSTM allows information to be safely stored for short or long periods and used only when necessary. The information looping functions as a sort of internal memory and may, in some cases, enhance the context visibility beyond what our history allows and up to an arbitrary depth. 
We add a small number of dedicated LSTM cells (with input, output and forget gates) on top of the existing network to examine whether this capability provides better predictability.

\subsection{Area and energy considerations}
\label{sec:feasibility}

\begin{figure}[t]
	\centering
	\includegraphics[width=0.85\columnwidth]{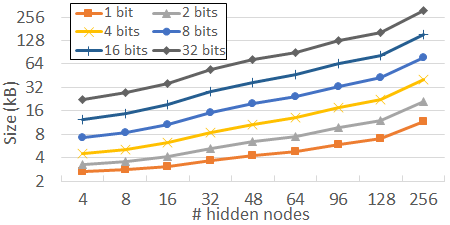}
	\caption{Storage required by the NN prefetcher based on size of the hidden layer and the precision bits (assuming 
	an association queue of 128 entries), log-log scaled. The circle shows the implementation used in this paper, which takes~\tilde{14kB}.}
	\label{fig:NN_storage}
\end{figure}

Implementing a neural network typically requires a significant die area and consumes a lot of power. This section describes how we mitigate the power and area overheads to achieve a feasible neural network prefetcher. Notably, power-performance efficiency may be lower than for other prefetchers, but given the rapid progress in NN research, we believe this can be further optimized in the near future.

The first mitigation, often employed in modern neural networks, is trading off precision for power and area. Common NN implementations employ half-precision floating point math (FP-16)~\cite{Movidius}, and some even reach 8-bit precision~\cite{TPU}, with a relatively small impact on the overall accuracy of the learning process. Such a reduction was shown to save area by up to 62\% per 2x bit reduction~\cite{FP16_FMA}, due to the number of operations and the simplification of the carry chains. Our implementation therefore uses 8-bit precision for FP calculations.

The fully connected topology of our 3-level neural network will include the following multiply-and-accumulate operations for the feed-forward operation (and approximately the same for the back-propagation step):
\begin{equation}
N_{fma} = N_{input} * N_{hidden} + N_{hidden} * N_{output} \approx 5k
\end{equation}

Based on estimations by Brunie~\cite{FP16_FMA}, a single precision FP FMA with fixed point accumulator would require \tilde 2000 $\mu m^2$ per cell on a 28nm process (adding mixed precision does not seem to add much in area, while improving precision significantly). On a modern 14nm process this should shrink by 4x, so our 32$\times$32 systolic array would require about \tilde $0.5 mm^2$ (a small fraction of a modern core). Adding the control logic, data paths and accumulators incurs negligible area overhead compared to the matrix itself. The other significant area consumers are the weights matrix (with an 8-bit weight per FMA operation) and the association queue (with 128 entries of 128-bit state and 8-bit neuron value for each of the 32+32 neurons), resulting in 15kB of storage. 
Figure~\ref{fig:NN_storage} shows how the overall storage is affected by the size of the hidden layer and the number of precision bits used.

The next concern is energy consumption. Based on Horowitz et al.~\cite{Computing_energy}, with 32bit floating point (single precision) each FMA operation would consume roughly 4.6pJ on a 45nm process, but a 16-bit FP FMA would take only take 1.5pJ, 3$\times$ less energy. 
Process scaling provides a significant reduction. Bohr claims~\cite{14nm} a \tilde1.6$\times$ improvement in energy efficiency per generation on an Intel process; therefore a neural network on 14nm should be \tilde{4$\times$} more energy efficient. We can also assume that reducing the precision from 16-bit to 8-bit would reduce the power and energy by an additional \tilde3-4$\times$, so the overall energy per step would be \tilde700pJ.


Finally, recent work shows rapid progress with quantized neural networks, where the values (weights and activation values) are reduced to a few bits~\cite{QNN}, with binary neural networks being the extreme example~\cite{BNN}~\cite{BNN2}~\cite{XORnet}. In addition to reducing the area significantly, these techniques make it possible to simplify the calculation steps by orders of magnitude. 
With a BNN, for example, all FMA operations on binary values are reduced to simple bitwise logic (XNOR and popcount). Higher precision designs also show a significant improvement in area and energy cost. Jouppi et al.~\cite{TPU} quote 6$\times$ less energy and area for 8-bit quantized multiplications, as well as 13$\times$ less energy and 38$\times$ less area for 8-bit quantized additions. 
The main hurdle with quantized networks is that until recently they were applied only for inference. However, recent work by Courbariaux et al.~\cite{BNN_journal} and by Tang et al.~\cite{BNN_train} shows promising results in the training domain as well. 
We adapted some of these techniques in our work to reduce the NN prefetcher overheads even further, but were not yet able to match the performance (as shown in Section~\ref{sec:qnn_results}). We therefore select an 8-bit FP precision for our design, but future work may be able to reduce it.

\section{Sequence prediction with NNs}
\label{sec:case_study}
Before testing the full prefetcher design on real workloads, we first inspect several variants for training our neural network in order to observe how well they predict sequences of values. We start with the basic patterns in Table~\ref{table:patterns} as a benchmark for real memory access streams. We also add several sequences based on functions that represent patterns of various complexities, as a proxy for more complicated access streams: a (shifted) sine function, a polynomial function, a linear line and a pseudo-random function (LFSR based).
Each series is fed into the neural network a single value at a time, 
and the output is trained to provide the next sequential element. 
Given a series $x^n_{i=1}$, we teach the neural network to predict the values in one of the following association modes: 
function estimation ($n \rightarrow x_n$), next element prediction ($x_{n-1}\rightarrow x_n$), next element with history ($\{x_{n-2},x_{n-1}\}\rightarrow x_n$), or delta prediction ($(x_{n-1}-x_{n-2}) \rightarrow (x_{n}-x_{n-1})$).

We measure the behavior of all benchmark sequences, using the 4 learning modes and 5000 iteration phases with varying neural network sizes and structures. Results are shown in square error per element (sum over all output bits), averaged over all elements in the sequence. 

\begin{figure}[t]
	\centering
	\includegraphics[width=0.95\columnwidth]{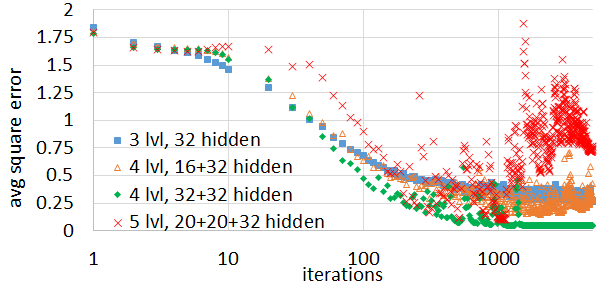}
	\caption{Average square error convergence over time when training the $sin(x)$ function over different NN depths and sizes (lower is better).}
	\label{fig:sin_conv}
\end{figure}

\begin{figure}[t]
	\centering
	\includegraphics[width=0.95\columnwidth]{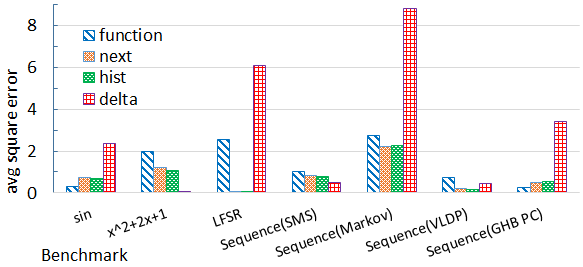}
	\caption{Average square error for different sequences and learning modes, on a network with 32 hidden nodes.
	}
	\label{fig:func_mode}
\end{figure}
\begin{figure}[t]
\centering
\includegraphics[width=0.95\columnwidth]{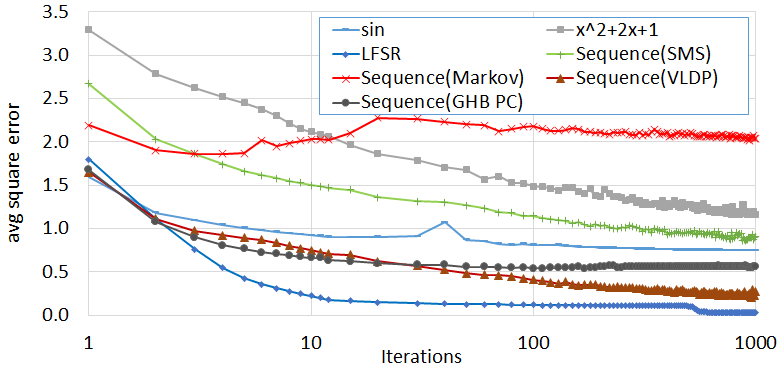}
\caption{Convergence rate for different benchmarking sequences.}
\label{fig:func_conv}
\end{figure}


Figure~\ref{fig:sin_conv} shows the average square error convergence over time for the same function, during the training process (sum of squared error per bit, averaged across all predicted values on a single cycle). 
As expected, the fastest converging network is initially the 3-level one, which reaches steady state in less than 1000 iterations. The two 4-level networks initially converge slower and are much more jittery during that process (as can be seen from the "cloud" of results) due to the interplay between the gradients of the two hidden levels, but eventually both surpass the 3-level network. The 5-level neural network fluctuates even more, as expected. However, it does not seem to converge on better results than the shallower networks, even when extending the process to $10^6$ iterations, which makes it impractical for online learning.

Figure~\ref{fig:func_mode} shows the average square error for the different benchmarks covered in this section. The figure shows the results for a 3-level network with 32 hidden layer neurons, after training over 5000 iterations (each covering all the elements in the sequence). We observe that different functions benefit from different correlation methods: the polynomial function, for example, benefits from delta learning as it reduces the polynomial degree. SMS benefits from deltas as well, since the pattern is intended to stress such recurring deltas. Markov and VLDP sequences, on the other hand, benefit from employing history in the input since this is the main way to distinguish between the Markov states. LFSR converges to an almost perfect prediction with any input that includes the previous element, due to the simple shift relations between elements (requiring that only the entropy bit be learned). Thus, for best coverage of real sequences, we need to implement multiple modes of correlation in parallel, and dynamically alternate between them. 

Figure~\ref{fig:func_conv} plots the rate of convergence over time for all the benchmark functions (using history learning mode as an example), over the first 1000 steps. The rate of convergence is relatively fast: most of the patterns had to be replayed only 100-200 times for the network to reach close to the final values, and the error remains stable beyond that point. 
The only slow-to-converge sequence is the Markov series, due to the order of the benchmark, which traverses each edge on the address "graph" several consecutive times to give it the desired probability before traversing the other edges.


\section{Related work}
\label{sec:related}


\subsection{Prefetching techniques}

Falsafi and Wenisch classified~\cite{PrimerHWPref} prefetchers into 3 groups: 
\begin{itemize}
\item \textbf{Stream/stride prefetchers} utilize \textit{spatial locality}, common in many applications that use linear data structures placed sequentially in memory. These prefetchers detect the constant stride pattern and run ahead of the demand stream. 
Most modern CPUs employ flavors of this family.
Recent work by Pugsly et al.~\cite{sandbox} proposed the Sandbox prefetcher, which tests different strides before choosing the optimal. The Best-Offset prefetcher by Michaud~\cite{Best_Offset}, which is based on a similar concept, has won the 2nd Data Prefetching Competition (DPC2)~\cite{DPC2}. Another participant in DPC2 was the Access Map Pattern Matching (AMPM) prefetcher by Ishii et al.~\cite{AMPM}, which detects the stride using shifted access pattern matching.

\item \textbf{Address-correlating prefetchers} utilize \textit{temporal locality} between pairs or sequences of accesses, indicating that accesses that appeared with some temporal adjacency in the past will manifest this adjacency in the future as well. The challenge is to isolate the correlated accesses out of a stream of unrelated ones. 
One of the original examples was the Markov predictor~\cite{MarkovPredictors}. Later work improved the prefetching depth, including the Global History Buffer Address-Correlation flavors (GHB/AC) by Nesbit and Smith~\cite{GHB}, which observe a long history of accesses and isolates recurrences from it (sometimes using the program counter for localization), and the Irregular Stream Buffer (ISB) by Jain and Lin~\cite{ISB}, which attempts to restructure the dataset spatially, similar to the instruction-based counterpart, the trace cache~\cite{trace_cache}. This category also includes works targeting linked data structures, such as those by Roth, Moshovos and Sohi~\cite{Roth98},~\cite{Roth99}, and by Bekerman et al.~\cite{Bekerman99}. These prefetchers track recurring memory accesses and generate jump pointers into irregular data structures. 

\item \textbf{Spatially-correlated prefetchers} use an extension of  temporal locality that correlates between spatial patterns instead of absolute addresses. These prefetchers seek out recurring spatial patterns that are not part of a long consecutive sequence but may repeat locally, such as accesses to the same fields of a structure across different instances. Examples of this family are Spatial Memory Streaming (SMS) by Somogyi et al.~\cite{SMS}, the DC flavors of GHB~\cite{GHB}, and the Variable Length Delta Prefetcher (VLDP) by Shevgoor et al.~\cite{VLDP}.
\end{itemize}

Semantic prefetchers can be seen as a combination of all the above. The wide context they use allows the correlation learning engine to extract address or delta correlation artifacts, or even other forms of relations between history elements. Such a technique was recently presented by Peled et al.~\cite{context_pref}, using a  \textit{contextual-bandits} scheme and a slew of hardware and software attributes.
However, that approach had to rely on complicated context hashing and dimension reduction mechanisms. 


\subsection{Neural network based predictors}
Using neural networks as a means to optimize micro-architectural speculations and predictions has already been proposed. Jim{\'e}nez and Lin proposed using perceptron-based neural networks for branch prediction~\cite{perceptron_branch_pred}. More recently, Teran, Wang and Jimenez extended this concept for predicting reuse distance and replacement policy~\cite{perceptron_reuse}. The CPU industry is also incorporating these techniques, with both AMD and Samsung publicly claiming to incorporate neural networks into their branch predictiors (in ``Ryzen"~\cite{AMD_nn} and in ``M1"~\cite{Samsung_nn}).

While predicting addresses adds a degree of complexity compared to binary decisions (taken vs. not-taken branch, or keep vs. replace), this work sheds some light on the feasibility of implementing a simple neural network over hardware. 

Due to the approximated results of neural networks, they seem to fit micro-architectural speculations where mistakes do not have functional effects. However, neural networks are not restricted to that domain. Some researchers also apply them to predict functionally visible results.
Esmaeilzadeh et al. presented a neural network based predictor for estimating function results~\cite{NNApproxFunction_Ceze}, used when some degree of approximation is allowed (such as some image processing kernels). Outside the CPU architecture domain, Knoll and Freitas~\cite{PAQ8} reviewed neural network models with \textit{stochastic memoization} for sequence prediction used to optimize compression algorithms.

Siegelmann and Sontag have shown recurrent neural networks (RNN) to be Turing complete, i.e., to have the computational complexity of a Turing machine~\cite{TuringComplete}, even with the constraint of having rational weights, thus making it possible to model them with real hardware.

Google teams proposed using neural networks as a Turing-compatible model of computation in some scenarios: Graves, Wayne and Danihelka~\cite{NeuralTuring} proposed training such a network to learn the actions of an actual Turing machine, allowing it to learn simple algorithms, while Kurach, Andrychowicz and Sutskever~\cite{NeuralRandomAccess} proposed using a neural network with controlled memory nodes (LSTM) to perform complicated algorithms involving linked data structure and array manipulations, focusing on the detection of memory access sequences of specific tasks with small footprints. 
Recently, Graves et al. published a complete neural network based compute model called \textit{differentiable neural computer (DNC)}~\cite{DNC}, augmenting the neural network with a novel form of internal memory based on parallel weighted read/write matrix operations. DNC was also shown to successfully learn short graph traversals and predict resulting nodes and shortest paths.
More recently, Hashemi et al.~\cite{Learning_Patterns_LSTM} presented an initial exploration of LSTM neural networks for memory access pattern prediction, relying on PC and address deltas as features. While they have not yet suggested a practical prefetcher, their work offers insights on the potential of deeper networks.

The computational and storage costs of neural networks remain critical limitations. Recent work attempts to mitigate these costs by employing dedicated eDRAM and optical interconnects for storing the weights (DaDianNao by Luo et al.~\cite{DaDianNao}), or by using resistive-memory based crossbars for storage and analog computation (Shafiee et al.~\cite{memristorNN}).
Other studies seek to simplify the cost of maintaining the NN node weights by reducing them to quantized (e.g. binary or ternary) values that will be easier to manage in hardware~\cite{BNN}~\cite{TNN}. This approach is mostly used for inference, but quantized training is also explored~\cite{QNN}.

\hide{
\subsection{Complexity limits}
Balc{\'a}zar and Gavald{\`a}~\cite{Kolmogorov} proposed a set of complexity classes describing recursive sequences (they used the complexity theory equivalent of \textit{languages} to the same effect), and have shown how the complexity is related to the amount of information the network can store in its weights, and at what precision (integer, rational or unbounded real numbers). 
}
\section{Methodology}
\label{sec:methodology}

\begin{table}[t]
	\scriptsize
	\centering
	\begin{tabular}{| l | l |}
		\hline
		Simulation mode & Sys. emulation, accurate timing, x86\\
		\hline
		Core type & OoO, 4-wide fetch \\    
		\hline
		Queue sizes & 192 ROB, 64 IQ, 256 PRF, 32 LQ/SQ \\
		\hline
		MSHRs & L1: 4, L2: 20\\
		\hline
		L1 cache & 64kB Data, 32kB Code, \\
		& 8 ways, 2 cycles access, private\\
		\hline
		L2 cache & 2MB per core, 16 ways, 20 cycles access, shared\\
		\hline
		Main memory & 2GB, 300 cycles access\\
		\hline
		\multicolumn{2}{c}{NN prefetcher} \\
		\hline
		Levels & 3, 4 or 5\\
		\hline
		Neurons & 128 input, 32 output, \\
		& 32-128 hidden neurons per layer \\
		& + 8 LSTM nodes (when applicable)\\
		\hline
		Association queue & \HistQ entries\\
		\hline
		
		\multicolumn{2}{c}{Competing prefetchers} \\
		\hline
		GHB (all)~\cite{GHB} & GHB size: 256, History length: 3 \\
		& Prefetch degree: 2, Overall size: 4kB \\
		\hline
		SMS~\cite{SMS} & PHT size: 2K, AGT size: 32, Filter Table: 32 \\
		& Regions size: 2kB, Overall size: 20kB \\
		\hline
		VLDP~\cite{VLDP} & 3 DPTs $\times$ 64 entries\\
		\hline
		Context RL~\cite{context_pref} & CST size: 2K entries x 4 links (18kB), \\
		& Reducer: 16K entries (12kB) \\
		& Prefetch queue: 128 entries \\
		\hline
	\end{tabular}
	\vspace{3pt}
	\caption{Simulator parameters.}
	\label{table:params}
\end{table}

\begin{table}[t]
	\scriptsize
	\centering
	\begin{tabular}{| l | l |}
		\hline
		BFS & Breadth-first search over a spatial \\
		& (array-based) 10k vertice graph\\
		\hline
		listsort & Linked list sorted insertion \& deletion \\  
		& (over 1M elements) \\    
		\hline
		matmul & Matrix multiplication (1k $\times$ 512 by 512 $\times$ 1k)\\
		\hline
		suffixArray & Create a suffix array of a 34M long DNA sequence\\    
		\hline
		setCover & Find a greedy approximate minimal set cover \\ 
		& over a 20k rMat based array represented graph \\
		\hline
		convexHull & Find a convex hull of a 500k 2D plummer map\\
		\hline
		spmv & Sparse matrix multiplication \\    
		\hline
		array & Sum a 1M long array\\    
		\hline
		bst & Binary search along a 10k array-based tree\\    
		\hline
		prim & Find a Prim min spanning tree over a linked \\
		& graph ($2^{12}$ nodes, rMat generated)\\
		\hline
		list & Pointer chasing along a 200k long linked list\\    
		\hline
		kruskal & Find a Kruskal min spanning tree over an \\ 
		&array-based graph (100k nodes rMat generated) \\    
		\hline
		patterns & Access along a 200k-long array using \\ & arithmetically growing index jumps\\    
		\hline
	\end{tabular}
	\vspace{3pt}
	\caption{Manual kernels.}
	\label{table:kernels}
	\vspace{-4mm}
\end{table}

The neural network based prefetcher was modeled on the gem5~\cite{gem5} simulator, using system emulation (SE) mode in order to focus on the application user-level code, and running an out-of-order x86 CPU for realistic behavior. Table~\ref{table:params} specifies the parameters of the simulated system. 

A large selection of frameworks is available for implementing neural networks (alexNet~\cite{alexnet}, Caffe~\cite{caffe}, torch~\cite{torch}). However, we preferred instead a simplified in-house model, both due to constraints in integration into our simulator, as well as to ensure that the implementation is feasible in hardware and has no hidden optimizations.

Our network parametrizes the depth (number of layers), the width (number of nodes per layer), and the connectivity (number of nodes in the previous layer connected to a single node in the current one). 
In addition, it allows adding an arbitrary number of nodes with memory connectivity and gates, modeling LSTM connections. 


We compare the NN-based prefetcher with four state-of-the-art prefetchers: VLDP~\cite{VLDP}, SMS~\cite{SMS}, GHB-PC/DC~\cite{GHB}, and Context-RL~\cite{context_pref}. VLDP represents a family of stride prefetchers and was shown to outperform other stride prefetchers (including the latest prefetching competition winner, Best Offset~\cite{Best_Offset}). The well-known SMS and GHB-PC/DC prefetchers represent temporal correlations prefetchers and pattern-based prefetchers. Context-RL uses a similar context-based prefetching concept but a simpler and limited learning mechanism.

Finally, we use a wide range of common benchmark suites, including SPEC 2006~\cite{spec2006}, Graph500~\cite{graph500}, and HPCS~\cite{bader05}.
We also add multiple hand-written kernels, to achieve high coverage of application behavior. Table~\ref{table:kernels} describes the manual kernels used. In this paper we focus on single-threaded applications to demonstrate how semantic locality prefetching can contribute even on that hard-to-scale class.

The benchmarks were compiled with an LLVM v3.6.2 compiler~\cite{LLVM}.
The simulation was done over 50M instruction phases, at skips of 50B/100B instructions selected according to memory workload characterization by Jaleel et al.~\cite{mem_workloads} to cover the main steady-state phases of memory behavior. For SPEC06 traces we usually captured 2-3 phases per trace, except when the hotspot did not show benefit for any of the prefetchers checked (usually due to low memory activity phases). 

Multi-threaded applications were not handled in this paper as we aim to explore the limits of single-threaded performance, which is far more limited by memory latency. 
MT runs may gain additional performance benefits from shared learning on similar threads, but this is left for future work.

\section{Evaluation}
\label{sec:evaluation}

\begin{figure}[t]
	\includegraphics[width=\columnwidth]{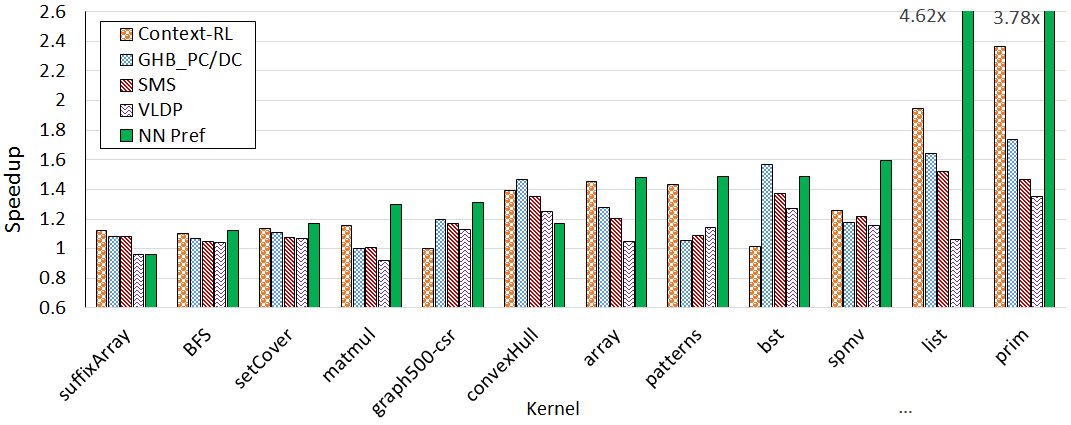}
	\\~
	\includegraphics[width=\columnwidth]{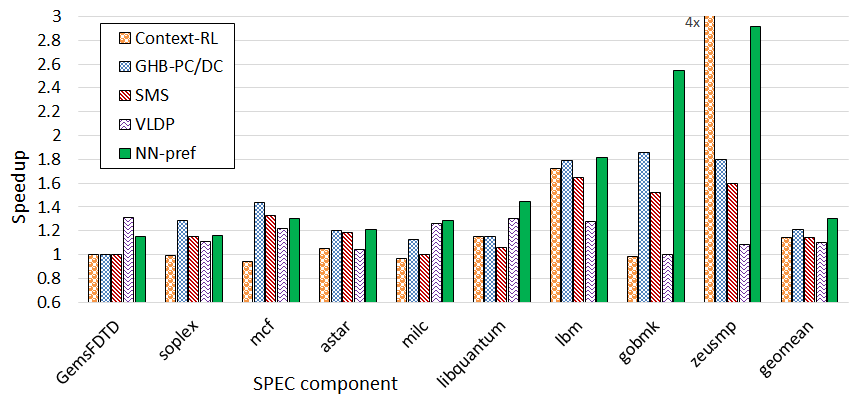}
	\caption{Speedup gain from NN based prefetching vs. other techniques, over a variety of applications from different benchmark suites. Only tests with gain above 5\% over any prefetcher are shown (SPEC suite includes c/c++ based benchmarks that could be compiled over LLVM).}
	\label{fig:speedup}
\end{figure}

\hide{
\begin{figure}[t]
\centering
\includegraphics[width=0.95\columnwidth]{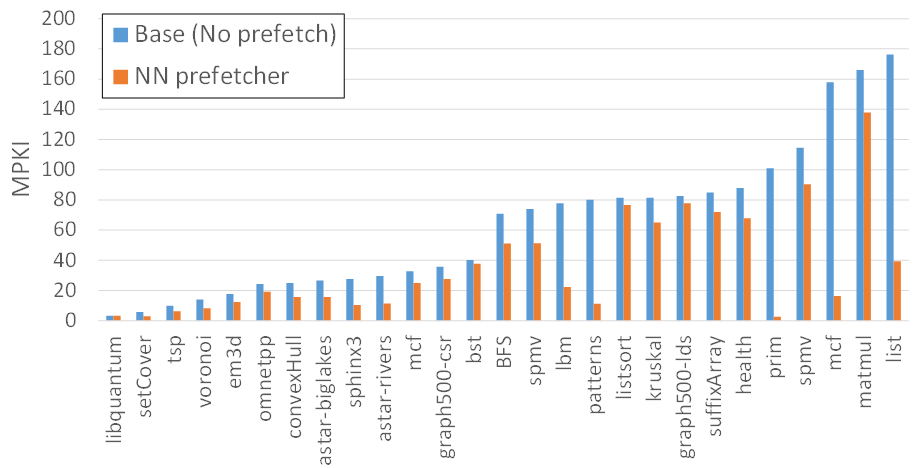}
\caption{Misses per 1000 instructions with the neural prefetcher vs. a baseline of no prefetching.}
\label{fig:mpki}
\end{figure}
}

The neural networks evaluated for our prefetcher have 3 levels (1 hidden layer with 128 neurons), 4 levels (2 hidden layers with 96 and 64 neurons respectively) and 5 levels (3 hidden layers with 96, 64 and 64 neurons). We also ran the 3 level network with an additional 16 LSTM nodes, making it a recurrent NN. 
Deeper networks are not tested here as even 5-level networks were shown in Section~\ref{sec:case_study} to converge too slowly. Convolutional neural networks (CNN) were also tested but are not shown since the results were not stable enough and the partial connectivity we used degraded their performance without exposing the desired spatial locality across input nodes.

Figure~\ref{fig:speedup} compares the IPC speedup of the 3-level NN prefetcher against other state-of-the-art prefetchers. The speedup is shown compared to a baseline with no prefetching. The workloads shown are the ones that exhibited some minimal degree of sensitivity to prefetching, getting a speedup of at least 5\% on any of the tested prefetchers. However, the geomean shown for SPEC2006 is over the entire suite, except for the Fortran benchmarks, which could not be compiled on our LLVM/clang, and some benchmarks that had build issues (gcc, perlbench) or were not supported by gem5.
The graph shows that most of the benchmarks with complex memory access patterns benefit significantly from the neural network based prefetchers. On average, the NN-prefetcher gains \AvgSPECGain, which is \SPECGainOverGHB higher than GHB PC/DC, \SPECGainOverSMS higher than SMS, \SPECGainOverVLDP higher than VLDP and \SPECGainOverRL higher than the RL-based context prefetcher. The most notable gain on SPEC2006 was in LBM (which averaged to 90\%, although some of the phases reached a local speedup of 2.8$\times$). 
On kernels, the benefits are even higher, with the max gain (on a linked list traversal test) above 5$\times$. 


\begin{figure}[t]
	\centering
	\includegraphics[width=0.95\columnwidth]{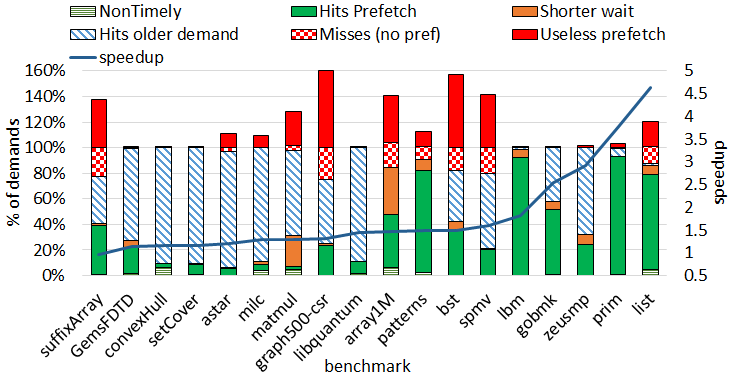}
	\caption{Breakdown of demand accesses according to the prefetch usefulness. Useless prefetches are stacked on top of the 100\% demands. Only tests with over 10\% demand L1 miss rate or with a large useless prefetch bucket are shown. The secondary y-axis} shows the speedup curve.
	\label{fig:miss_buckets}
	\vspace{-3mm}
\end{figure}
The performance gain of a prefetcher depends on the tradeoffs between useful, useless, and partially-useful prefetches.
Figure~\ref{fig:miss_buckets} shows the breakdown of demand misses in the L1 cache. Each of them can be categorized as one of the following (sorted by increasing usefulness): a miss that was never prefetched, a prefetch that was triggered but not yet sent (non-timely), or a prefetch that was sent but not yet completed (shorter wait). In addition, we have demands that hit cached lines already hit by previous demands, and useful prefetches (only the first demand to hit them). 
On top of these categories covering 100\% demands, we add bad prefetches: addresses prefetched but never used while in the cache, indicating that the prefetcher was wrong about the address. Speedup is correlated with having many useful prefetches but not too many useless ones. In some cases (e.g., matmul), most of the gain is from reducing the demand miss latency, indicating that the prefetcher is successful, but further depth tuning may provide better results.

\subsection{Comparing different NN schemes}
\begin{figure}[t]
	\centering
	\includegraphics[width=0.85\linewidth]{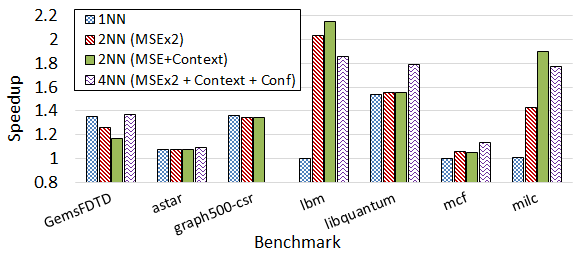}
	\caption{Comparison between different association policies for training address selection. \textit{MSE} chooses for each output subset the address closest to the network output, \textit{Context} policy prefers matches in the context hash, and \textit{Conf} prefers associating addresses of frequent PCs.}
	\vspace{2pt}
	\label{fig:parallel_nn}
\end{figure}
\begin{figure}[t]
	\centering
	\includegraphics[width=0.88\linewidth]{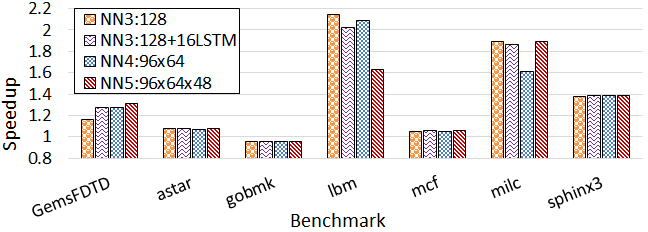}
	\caption{Comparison between NN sizes, depths, and LSTM support.}
	\label{fig:nn_levels}
	\vspace{-3mm}
\end{figure}

Figure~\ref{fig:parallel_nn} shows the performance of the different association schemes described in Section~\ref{sec:nnpref} (over a single phase at 50G instructions skip). The NN is partitioned and trained over different association candidates, producing multiple prefetch candidates. 
We observed that 4 parallel networks are not necessarily better than 2, due to additional thrashing (GemsFDTD, for example, gains less when we activate more networks), and because the PC-based association is less effective (LBM and milc). The remainder of the results were obtained using the 2NN scheme. 

Figure~\ref{fig:nn_levels} shows the impact of network size (number of neurons in the hidden layers), depth, and the use of special features such as LSTM nodes. Interestingly, the differences are very small, and some benchmarks exhibit opposite trends (LBM prefers smaller, simpler networks, while GemsFDTD and MCF prefer larger networks). These negligible differences concur with the observation that deeper networks cannot perform online training fast enough against changing code phases to extract any benefits from their size. 

LBM benefits the most from a simple neural network while MCF, GemsFDTD and astar benefit slightly from adding LSTM nodes. This is due to the reuse distance of pattern types: MCF cannot memorize its full linked data structures as they are too big. It may therefore gain from long term memory by storing some critical addresses in the LSTM neurons (such as those in the first few levels of its cost network). The same applies for other applications using linked data structures such as astar. LBM, on the other hand, has a grid layout that is more spatially recurrent and can train a generic pattern that does not require long term memory.

\subsection{Classifying prefetch usefulness}
Different applications demonstrate different classifications for prefetcher usefulness. 
First, there are the spatially organized applications, where the program traverses data structures sequentially. In these cases, the neural network prefetcher can learn a constant delta, but so can the other prefetchers examined. 
The neural network prefetcher exhibits higher gains thanks to better context localization (from which the GHB-PC/DC also benefits, albeit to a lesser extent). In some of the cases, the neural network prefetcher also gained thanks to the fixed distance policy (associations are always made at a fixed history distance), which improved its coverage.

The second category is temporally correlated applications such as linked lists. These cases are almost impossible to describe as a function, as they exhibit an almost random dataset layout. The recurrence can, however, be recorded and replayed, at least up to the length that the prefetcher storage can support. 
In this category (and most notably in the link-based version of SSCA), the context-RL prefetcher often wins due to the speed of its learning (a single access is enough to generate a complete prediction). However, on larger data-set sizes, the neural network will be able to scale better than the context-RL prefetcher thanks to its distributed storage.

\subsection{Comparison with other prefetchers}

The neural network prefetcher and the context-based RL prefetcher both learn by associating 
between context states and addresses. Comparing them will therefore illustrate the difference between the storage and complexity limitations (although there may also be minor differences in the policies used by the two methods to select the best associations).

For some of the applications we can see that context-RL provides a higher speedup, the most notable example being the list-sort kernel. 
The results show that the list itself is too big to fit in any of the prefetchers in full (allowing only a subset of the list to be stored). Because the sorted list is built gradually, a limited head segment of the list can easily fit since it has a better chance to be used during every search. 
However, this segment may frequently change while elements are being added with uniform distribution across the list. The context-RL prefetcher can update the CST table on such changes with a better association (a process that occurs instantly once the score of the new association exceeds that of the old one). Conversely, the neural network prefetcher has to rebuild its weights to reflect the change. Since each of these weights represents multiple elements in parallel, this process may break other predictions and may take a long while to reconverge every time the data structure changes. 

In some cases, however, applications such as LBM and kernels such as prim and matmul show significant gains compared to the context-RL prefetcher. These gains are attributed to better storage and due to contextual associations (mostly on a complex pattern, as in prim and LBM), and partly due to association policies that favor constant lookahead depth (which is useful for spatial traversals, as in matmul).

Gobmk shows an interesting gain unique to NN-prefetch (only on the longer skips). This Go benchmark recursively plays game moves down the decision tree and recovers them when they are pruned. This makes the stack of board states big, but the allocation scheme groups related moves spatially. The descent depends on prior decisions, making the branch/PC history an effective heuristic for the chosen paths. The NN usefulness is only observable on the longer skips, when the game stack becomes larger and more fractured.

\subsection{NN quantization}
\label{sec:qnn_results}
\begin{figure}[t]
	\centering
	\includegraphics[width=0.95\columnwidth]{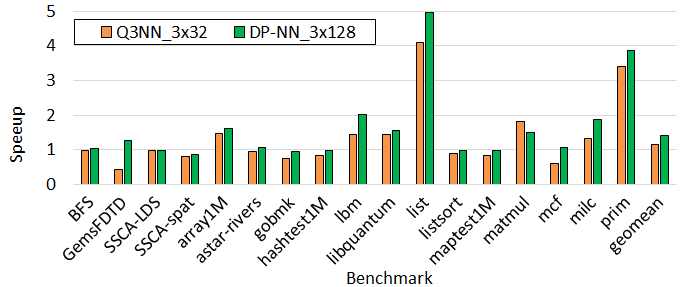}
	\caption{Comparing a network using double-precision FP and 3-bit QNN.}
	\label{fig:qnn}
	\vspace{-6mm}
\end{figure}
Figure~\ref{fig:qnn} shows a comparison between a 3-layer network (with 128 hidden nodes) that uses 32-bit floating-point weights and activations and one that uses 3-bit quantized values (the input and output values are always binary). Most benchmarks sustain only a small hit in their speedup, with the exception of MCF and gemsFDTD, which incurred significant slowdown due to their inability to converge with the quantized gradients.

\section{Conclusions}
\label{sec:conclusions}

This paper presents a neural network memory prefetcher based on \emph{semantic locality}. This recently proposed model argues that locality of reference is an artifact intrinsic to the program implementation rather than a simple \emph{spatio-temporal} correlation.
The proposed prefetcher learns the algorithmic properties of programs by feeding machine and program state elements as inputs to a neural network. The NN is trained at runtime to predict future memory accesses based on correlated context-address associations.

Our main goal in this paper is to examine whether the learning capabilities of a neural network are inherently superior to those of other machine learning or heuristic-based techniques. 
The NN prefetcher covers a spectrum of spatio-temporal access patterns that can only be handled today by multiple heuristic \emph{spatio-temporal} prefetchers working together. Our analysis demonstrates that this prefetcher outperforms other state-of-the-art prefetchers, providing~\SPECGainOverGHB higher gain on SPEC06 than GHB-PC/DC,~\SPECGainOverSMS more than SMS and \SPECGainOverVLDP over VLDP.

Our current design, while still expensive in terms of area and energy efficiency, exceeds state-of-the-art prefetchers in performance but not necessarily in power/performance efficiency. As such it may serve well in high-power designs. However, the fast evolution of NN technology will make future networks faster, more compact, and more efficient, thereby making the NN prefetcher more accurate and power efficient. 



\bibliographystyle{ieeetr}
\bibliography{ref}

\end{document}